# PAL: A Wearable Platform for Real-time, Personalized and Context-Aware Health and Cognition Support

**Mina Khan**     **Glenn Fernandes**     **Utkarsh Sarawgi**     **Prudhvi Rampey**     **Pattie Maes**

MIT Media Lab
Cambridge, USA

{minakhan01, glennf2802, usarawgi911, prudhvi.rampey} @gmail.com, pattie@media.mit.edu

## ABSTRACT

Personalized Active Learner (PAL) is a wearable system for real-time, personalized, and context-aware health and cognition support. PAL's system consists of a wearable device, mobile app, cloud database, data visualization web app, and machine learning server. PAL's wearable device uses multi-modal sensors (camera, microphone, heart-rate) with on-device machine learning and open-ear audio output to provide real-time and context-aware cognitive, behavioral and psychological interventions. PAL also allows users to track the long-term correlations between their activities and physiological states to make well-informed lifestyle decisions. In this paper, we present and open-source PAL's system so that people can use it for health and cognition support applications. We also open-source three fully-developed example applications using PAL for face-based memory augmentation, contextual language learning, and heart-rate-based psychological support. PAL's flexible, modular and extensible platform combines trends in data-driven medicine, mobile psychology, and cognitive enhancement to support data-driven and empowering health and cognition applications.

## INTRODUCTION

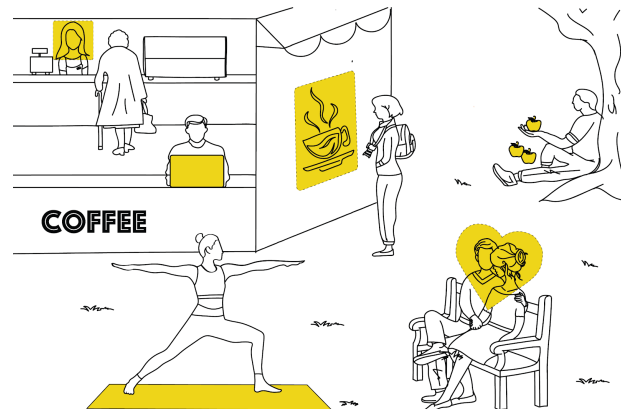

**Fig 1. PAL's vision of health and cognition support**

One in four people in the world [33] and one in five people in the United States [34] are affected by mental health problems. Over 70% deaths globally [3] and about 88% in America [4] are due to non-communicable diseases (NCDs) like cardiovascular disease, cancer, and diabetes, most of which are preventable. Recent advancements in Artificial Intelligence (AI), big data, and ubiquitous computing allow us to support human health and cognition in at least 3 ways:
i. Data-driven medicine, holistic P4 (predictive, preventive, personalized and participatory) medicine using big data, systems biology, and personalized lifestyle design [8];
ii. Mobile psychology, i.e. mobile and wearable for psychological help [18] and behavior change support [1];
iii. Enhanced cognition or Intelligence Augmentation (IA), i.e. coupling human and machine intelligence to support human memory, learning, decision-making, etc [24].

We believe that effective and holistic health and cognition support requires combining data-driven medicine, mobile psychology, and enhanced cognition to provide real-time, context-aware and personalized support to users. While data-driven medicine can give actionable insights to the users, mobile psychology and cognitive enhancement can provide behavioral, psychological and cognitive support to empower the users to act on personalized insights.

Personalized Active Learner (PAL) is a wearable system that marries data-driven medicine, mobile psychology, and enhanced cognition to provide real-time, personalized and context-aware health and cognition support. PAL's wearable system allows users to track their internal physiological states and external activities in real-time and over time to support self-awareness. Also, PAL's holistic user tracking, on-device machine learning, and wearable feedback allow for real-time, personalized, and context-aware health and cognition support. Using PAL, we envision a future where users are empowered with actionable insights and real-time action support to improve their health and cognition (Fig 1).

In this paper, we present the design, implementation, and example use cases of PAL's flexible, modular, and extensible platform for holistic user tracking, machine learning, and personalized support. PAL's open-source platform aims to enable designers, developers, and researchers to do rapid prototyping of real-time, personalized, and context-aware cognitive, behavioral and psychological support applications. Medical professionals can also use PAL to get personalized, detailed, and real-time information about their patients to help them in personalized and real-time ways. People can also contribute the behavioral and physiological data they gather using PAL to open databases to improve health and cognition research.

## RELATED WORK

### Data-driven Medicine

There are several types of wearable sensors, from activity monitors to biosensors [15]. There has been research on using wearables for recording user context, e.g. Sensecam [6], a lifelogging camera, and also on multi-modal sensors for tracking human activities in relation their physiological states, e.g. SenseWear [23], a platform for multimodal physiological sensing, eButton [28], a wearable camera for health monitoring, and Startlecam [23], a Galvanic Skin Response-triggered camera. The Quantified Self movement [6] is evidence that users are interested in self-tracking and research shows that personalized health insights can improve individual health [21]. PAL combines activity and physiological sensing with on-device deep learning to track and correlate user's activities and physiological states in real-time and over time to provide personalized support.

### Mobile Psychology

There several examples of mobile psychotherapy, e.g. self-directed therapy [2], and cognitive behavioral therapy (CBT), e.g. for drug addiction and PTSD [7], and insomnia [16]. There are also behavior change technologies, including persuasive technologies and behavior intervention technologies (BITs) [1,19]. Behavior change technologies employ a wide range of behavior change techniques (BCTs) [17], including mindfulness-based interventions (MBIs) [20] and motivational interviewing [27]. Research shows that integrating behavioral and contextual tracking in mobile psychology applications can improve health, e.g. social well-being [13] and reduced marijuana use [27]. PAL supports flexible, personalized and context-aware feedback for psychological interventions and behavioral support.

### Enhanced Cognition

There has been significant work on wearables for cognitive enhancement, especially camera-based wearables, e.g. Memory Glasses for memory support [31], Sixthsense for real-time information delivery [31], SocioGlass for face recognition [32], and FingerReader for assistive visual interpretation [26]. There are also non-camera-based feedback systems, e.g. for real-time attention [14] and emotion regulation [5]. Most cognitive enhancement wearables, however, serve a single purpose and do not track user's holistic activities and physiological states to flexibly assist users in a range of context-aware scenarios. PAL's context-aware wearable can adaptively assist users in different cognitive and psychological scenarios, from memory augmentation to physiological regulation.

PAL leverages trends in data-driven medicine, mobile psychology, and cognitive enhancement to facilitate holistic user tracking and real-time, context-aware, and personalized behavioral, psychological and cognitive support.

## DESIGN

We wanted to make PAL a platform for holistic tracking of user activities and physiological states to provide real-time and over time self-awareness and cognitive, behavioral and psychological support. We decided to make PAL a wearable system so that it can be always present with the users. PAL has five components (Fig 2): i. Wearable device with on-device machine learning to monitor user context and give real-time feedback; ii. Mobile app to allow users to control and customize the wearable device; iii. Cloud database to store user data for long-term tracking; iv. Web app to allow users to visualize their data; v. Machine learning server for additional off-device machine learning on the data from the cloud database. While parts i)-ii) provide real-time support, iii)-v) are for long-term tracking.

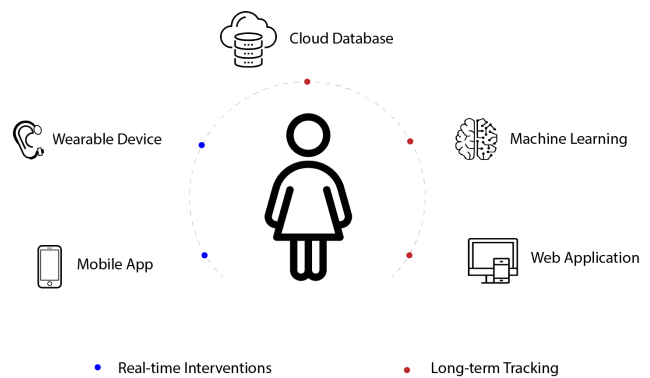

**Figure 2. PAL's system with five key components**

### A. Wearable Device

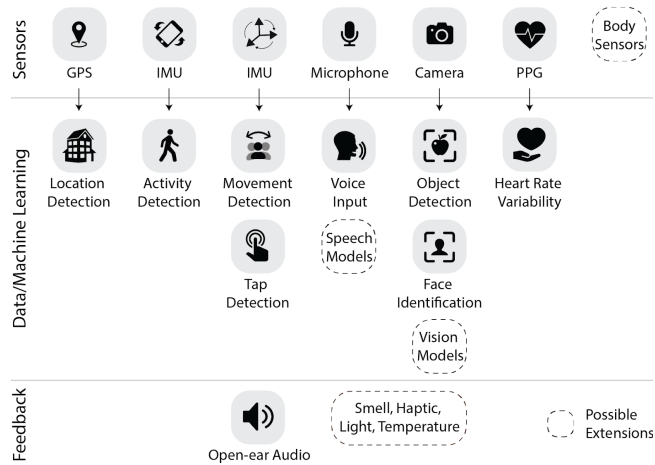

**Figure 3. PAL's wearable system components**

We included activity and physiological sensors, on-device machine learning, and real-time feedback on the wearable device to provide real-time and context-aware support (Fig 3). The wearable is modular and extensible, i.e. components can be removed or added, e.g. new Bluetooth-based sensors.

*Inputs and Sensors*
We selected the following five mobile sensors:
1. Geolocation and physical activity sensors: Geolocation and activity trackers are useful for digital phenotyping and activity recognition, and are available on most cell phones. We added geolocation and physical activity sensors to PAL due to their ubiquity and usefulness.
2. Camera: A wearable egocentric camera is good not just for Human Activity Recognition, but also for other augmentation applications, e.g., memory augmentation, etc. Advances in deep learning allow us to not only extract meaningful information from images but also to perform real-time on-device deep learning on images [11]. We added a camera in PAL to track user's activities.
3. Microphone: In addition to sensor data, we wanted direct user voice input. The user's voice can be useful for voice notes, voice interfaces, speech and emotion recognition, and audio messaging. Audio input can also record ambient sounds for activity recognition and environmental sound classification. With real-time audio recognition models for embedded devices, the opportunities for audio input on wearables are endless.
4. Heart Rate Sensor: We decided to include a physiological sensor that could be easily integrated into our wearable device and has high usefulness for the user's physiological state. We included a heart rate sensor to monitor heart rate variability (HRV), which is indicative of the user's sympathetic and vagal activity [22], and has applications for helping with stress, anxiety [9], etc.
5. Inertial Measurement Unit (IMU): Even though phones commonly have IMUs, we decided to include an IMU on the wearable device to detect physical movements of the user when they are not carrying their phones. IMU can be helpful for detecting physical activity, taking photos based on user movements, and cleaning up movement artifacts in sensor data, e.g. heart rate sensor. The IMU also has tap detection for direct user input.

*On-device Processing and Machine Learning (ML)*
Deep learning models are useful for processing sensor data, but deep learning often requires bulky Graphical Processing Units (GPUs). Machine learning accelerator chips allow for on-device inference of deep learning models, and we decided to add a machine learning accelerator chip to our wearable device to allow for real-time inference of deep learning models, e.g. for computer vision. On-device processing reduces the latency and power overhead of communicating with a remote device, and user data is also more private as sensitive information does not have to leave the local device if long-term records are not needed.

*Feedback/Output to the user*
Open-ear audio: There are different output modalities, e.g. haptic, audio, and visual. Open-ear audio can be seamlessly and simultaneously delivered without blocking the user's real-world content. Audio feedback has been shown to help behavior change, e,g, to promote exercise [10], and audio output can also be personalized for users, e.g. users can upload/select different sound clips or even record their own voice as audio reminders. We chose open-ear audio for PAL due to its seamlessness, effectiveness, and personalizability.

### B. Mobile App
We included a phone app to set up and customize PAL's wearable device, e.g. to select personalized audio output, connect the wearable to the internet, allow the user to log in with a unique ID for uploading data to a cloud database, turn on/off different sensors for custom applications, etc.

### C. Cloud Database
We added a cloud database for long-term tracking. The wearable device can be connected to the internet and each user's data can be periodically sent to a cloud database.

### D. Web App
We decided to use a web app to allow users to visualize and compare different datasets. Unlike a mobile app, users can use the web app on their computers as well as cell phones. Also, unlike mobile apps, web apps can easily use web visualization libraries to show creative visualizations. If needed, the mobile and web app can be merged int one app.

### E. Machine Learning Server
We decided to have a GPU-powered cloud server to perform machine learning on the data from the cloud database and then push the results back to the cloud. Since the machine learning accelerator can only infer deep learning models, the server can train/retrain deep learning models and also infer from non-mobile-optimized models.

## IMPLEMENTATION

We implemented the five components described in the design section. All code and component details are on Github: [removed for anonymity]

### A. Wearable Device

The wearable has ear-worn components connected to an on-body clip-on (Fig 4, 5). The components are as follows.

*Ear-worn Components*

1. 3D-printed ear hook: We designed five ear hook sizes using the average ear lengths and widths of different male and female age groups [12] ({Length,Width} = {58.5, 30}, {60.5,31}, {61.0, 31.5}, {63,32}, {65, 33}).
2. Camera: There are many places for placing the camera, e.g. belt, glasses frames, shirt, etc. We wanted the camera to view the same scene as the user, so we could not place it on a belt buckle or shirt because those positions are far from eyes. Putting a camera on glasses is common, but we did not necessarily need the full glasses frame on a user's face. We place a small (8.6mmx8.6mm) camera on an ear-hook such that the camera is positioned on the earlobe to optimally capture user's visual context.
3. Audio output: For open-ear audio output, we initially chose bone-conduction transducers (BCTs), but BCTs need good skin contact and a slight gap can cause a significant drop in audio quality. We found a small speaker (10mm x 15mm, 0.8g), and placed it near the ear opening on the ear hook for open-ear audio feedback.
4. Audio input: We added a small microphone (16.7mm x 12.7mm, 0.4g) on the wire going from the ear hook to the on-body clip-on. We tried placing the mic behind the ear but the audio quality was not optimal. By placing the mic on the wire near the user's face, we get good audio for the user's voice input. Also, since the microphone is not fixed, the user can bring it closer to their mouth to speak. We discuss the mic quality in the evaluations section.
5. Heart rate sensor: Electrocardiogram (ECG) is commonly used to calculate HRV, but most ECG devices require electrodes near the heart. For the sake of convenience and ease of everyday wearability, we did not want the user to have to put another device on their chest along with our ear-worn wearable. Previous research shows that photoplethysmography (PPG) can be a good estimate for ECG, and also PPG from the earlobe is comparable to ECG [29]. Thus, instead of using ECG near the chest, we added PPG on the earlobe to get HRV.

*On-body Clip-on Components*

1. Raspberry Pi Zero: Raspberry Pi Zero W gave us a good trade-off between weight/size, computational power, and power consumption. Also, we found the Raspberry Pi to be easier to program than its alternatives, e.g. NanoPi.
2. ML accelerator: For our machine learning accelerator, we chose the Intel Movidius Neural Computer Stick (NCS) because of its small size, compatibility with the Raspberry Pi Zero, and easy availability. Our Tensorflow models deployed on NCS are also compatible with the new NCS 2, and we may replace NCS with a newer ML chip, e.g. NCS 2 or Google's Edge Tensor Processing Units, both of which can run our Tensorflow models.
3. Accelerometer: We chose the ADXL345 chip as it supports activity/inactivity detection and tap detection.
4. Battery pack: We use a 3.7V 1500mAh Lithium Polymer battery. The battery size (50mmx30mm) is comparable to the Raspberry Pi Zero's size (65.0mm x 30.0mm) to make a compact enclosure for our wearable.
5. PowerBoost: We use a 5V PowerBoost to convert the 3.7V battery output to 5.2V power for the Raspberry Pi.

*Raspberry Pi software*

The Raspberry Pi side of the code is written in Node.js and Python, and it supports all the functionality of PAL's wearable device, e.g. taking pictures, giving audio output, recording audio using microphone, connecting the device to phone over Bluetooth, receiving information from phone over Bluetooth, sending data to cloud database over WiFi, running on-device deep learning models for object and face recognition, tap detection for detecting user tap input, activity/inactivity detection from the IMU, and doing heart rate computations from the PPG sensor. Node.js runs different Python operations simultaneously so different operations do not block each other. PAL performs on-device object detection using MobileNet-Single Shot Detector (SSD) model trained on 20 classes from the COCO dataset [11]. PAL also does on-device one-shot personalized person recognition using k-means clustering and FaceNet model [25]. All the PPG-HRV computations also happen on the Pi for real-time detection of user states.

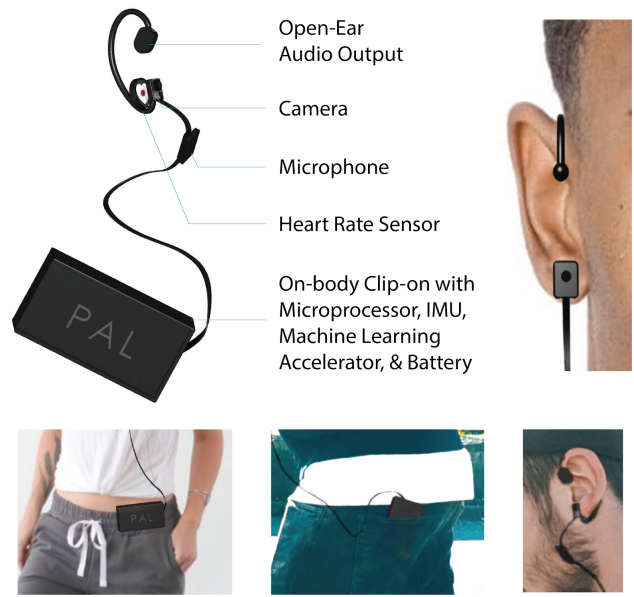

**Figure 4. PAL's wearable device with ear-worn parts connected to on-body clip-on components**

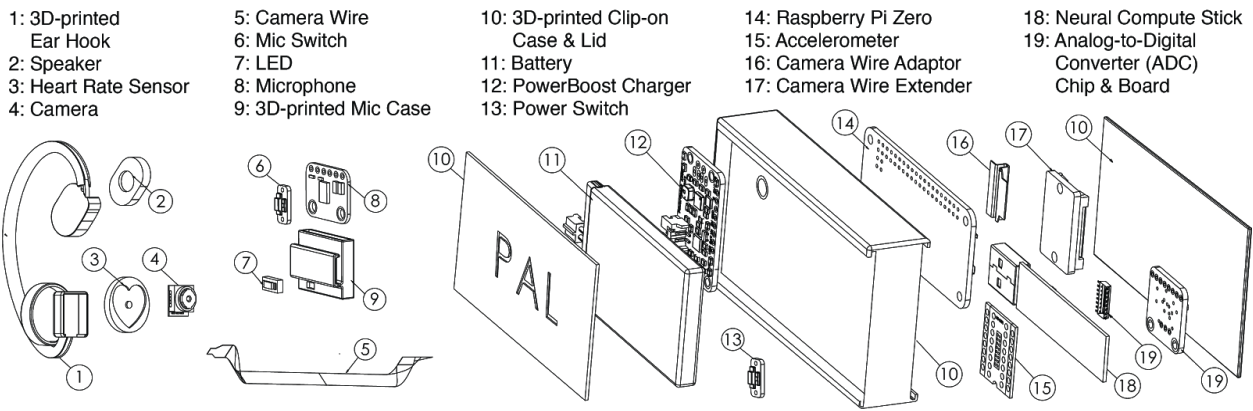

**Figure 5. Exploded view of the ear-hook (6cmx3cm), mic case (2cmx2cm) and clip-on hub (10x5x2cm)**

### B. Mobile App
We created an Android app to connect to the Raspberry Pi over Bluetooth. Our Android app uses Google's Location and Place API to get user's raw location (latitude and longitude), location name, and location type, and also Google's Activity Detection and Activity Transition API to get user's physical activity and activity transitions. Using PAL's mobile app, users can not only select/deselect particular sensors or detection models, but also select personalized audio for context-aware support, e.g. based on heart-rate, location changes, physical activity, face recognition, etc. The Android app has an optional Firebase Sign-In to allow us to send user data to the cloud if needed.

### C. Cloud Database
We opted for Firebase Database and Storage because of its low cost and ease of use. People can use their own database with PAL or setup up a Firebase one like us.

### D. Web App
We built a responsive web app using React.js, and used the same Firebase Sign-in in our web app as our Android app to get the same unique user identification for getting data from Firebase. Our website is deployed using Firebase, and we use custom visuals and D3.js library to visualize user data.

### E. Machine Learning Server
We set up a local server on a computer to run Tensorflow models on cloud data and to train deep learning models.

### EVALUATION
We conducted the following evaluations for our wearable.

### A. Camera View
With our ear-worn camera, our goal was to capture the user's visual context as close to the user's focus view as possible. Our camera has a 64° wide angle of view, which is comparable to the 60° wide cone of focus of a human eye. Since the angle of focus of a human eye is similar to our camera's angle of view, we placed one of our cameras in the middle of two eyes (to approximate human focus view) and another one on PAL's ear hook. Compared to the mid-eye camera, our right ear hook camera missed the left 200 cm and the top 100 cm of a 1200 cm x 750 cm view, capturing ~70% of the user's focus view for an image ~1 meter away.

### B. Wearability Study
We conducted a study (N=10; 5 males, 5 females; 4 aged 18-25, 4 aged 25-30, 2 aged 31- 40) to evaluate the comfort levels and sizes of PAL's wearable device. For each of the participants, we asked them 3 questions: Q1: What ear hook size fits you best? (5 sizes) ; Q2: How comfortable was the ear hook that fit you best?; Q3: How comfortable was the clip-on hub?. For Q2 and Q3, we evaluated comfort using a 5-point Likert scale, with 5 being very comfortable and 1 being very uncomfortable. All participants found that at least one of the sizes fit them well, and all sizes had at least one participant who picked it as their best fit size, which shows that our sizing had enough range to fit all our participants, but not too much range have redundant sizes. For ear hook comfort (Q2), 60% of the participants gave the ear hook a score of 4 or 40% a 5. The participants pointed out with the ear-hook, they did not feel like they were wearing anything, but some also added that it might change if they had to wear the ear hook for hours. In terms of the clip-on hub comfort (Q3), 50% of participants gave a score of 4 and 50% a 5. The participants found the clip-on comfortable and secure as did not fall off during walking.

### C. Audio Input/Output Study
We conducted another study (N=10; 5 males, 5 females; 3 aged 18-25, 5 aged 25-30, 2 aged 31- 40) to evaluate audio input and output quality with ambient sounds. We added 3 ambient sounds (traffic, office, rain), each at 3 different sound levels (50, 60, 75 dB). For both output and input quality evaluations, we tested 10 conditions each: 1 with no ambient sound; 3 with 50 dB sounds of rain, office, and traffic; 3 with 60 dB sounds of rain, traffic and office, and similarly 3 with 75 dB sounds. For testing audio output, we played 30 seconds of audio through PAL's open-ear audio output and for audio input test, we recorded 30 seconds of user's voice using PAL's microphone and then played it back to the user. For each of the 20 conditions, we asked the users how understandable the audio was on a 5-point Likert scale, with 5 being clearly understandable and 1 being not understandable at all. Our audio output was understandable with ambient sounds up to 50 dB, understandable with 60 dB noise, and not so understandable with 75 dB noise (avg. 5-point Likert scores: no added noise: 4.2; 50 dB noise: 3.6; 60 dB: 2.9; 75 dB: 2.2). We recommend using wordless sounds for loud environments. Our microphone input is clear even in loud environments (avg 5-point scores: no added sound: 4.8; 50 dB sound: 4; 60 dB: 3.2; 75 dB: 2.5).

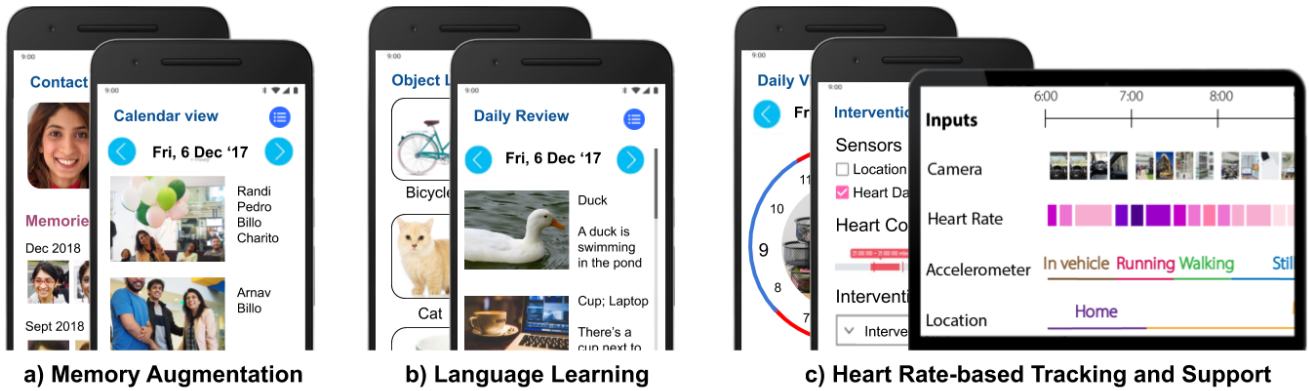

Figure 6. Mobile phone interface for three example applications built using PAL's platform

### D. In-the-wild Evaluation

We had 1 person wear the device continuously for 2 5-hour intervals to collect data in the wild using PAL. The wearable continuously ran heart rate, Bluetooth, and WiFi at 0.3A (idle Raspberry Pi consumes 0.2A at 5V) while the Android app collected activity and location transitions. After every 15 minutes or every activity/location transition, the wearable took a picture and performed object detection (<5s) and face recognition (<10s) at 0.45A. Our 1500mAh battery lasted 5 hours and we collected 50 pictures in which 33 of the 41 faces instances were recognized correctly and 78 of the 98 actual object instances were detected correctly.

### E. Heart-rate Evaluation

In order to compare the accuracy of our PPG-HRV, we compared it with Series 4 Apple Watch, Zephyr Biopatch ECG, and Empatica E4's PPG. Our HRV algorithm does wavelet denoising, trend removal, peak enhancement, and peak detection. The Apple Watch can only do 30-second recording, so we compared 10 30-second recordings each for 5 people wearing PAL's PPG sensor and Apple Watch. We found an average correlation of 91.2% between HRV from our PPG sensor and Apple Watch's ECG. We also compared 5 30-minute sitting datasets for 1 person wearing PAL's PPG sensor, Empatica E4's PPG, and Zephyr ECG. We got an average correlation of ~75% between HRV from our PPG and Zephyr ECG (Dataset 1-2: >90% correlation, and 3-5: 60-70%), which is more than the ~69% correlation between HRV from E4 PPG and Zephyr ECG. We found our PPG comparable to the Apple Watch, E4 and Zephyr.

## APPLICATIONS

We propose the following 3 example applications for health and cognition support using PAL. We did not test the applications but fully developed them to be quickly tested. Fig 6 shows the mobile interfaces for the three applications.

### A. Memory Augmentation via Face Recognition

PAL performs real-time personalized on-device face recognition to help people recognize known acquaintances. PAL can perform continuous face recognition (without user trigger) or on-demand face recognition (triggered by the user's tap on the device). Name of individuals or other relevant information is communicated to the user using open-ear audio. Users can add new faces to the database using just one image and the names of the people can be recorded using PAL's mic. Our cloud database and Android app allow people to record and review their contacts and people-specific memories over time as shown in Fig 6a.

### B. Contextual Language Learning in the Real-world

Using real-time on-device object detection, PAL recognizes objects in the real world and says their name in English for contextual English language learning. The images captured by the camera are sent to the cloud database from where our machine learning server pulls the images to run the Show and Tell image captioning model [30] to generate scene descriptions on each image. Scene descriptions are pushed back to the cloud database from where our mobile app pulls the images, object names and scene descriptions for the user. While the wearable device supports real-time learning, the mobile app is for long-term review (Fig 6b). Scene descriptions include more vocabulary than just object names and so we run the non-Movidius-compatible Show and Tell model on our server to generate scene descriptions.

### C. Heart Rate-based Awareness and Support

PAL allows the users to track their holistic day-long heart rate over time. As shown in Fig 6c, people can choose personalized audio interventions for a custom non-ideal HRV range. Also, PAL can visualize how their HRV was for the whole day and figure out the times/days when their HRV was non-ideal in real-time and over time. When the user is sitting, PAL continuously measures HRV and records user activity (visual context, location, and physical activity) at regular 15-minute intervals or after transitions, e.g. location or physical activity change. When the HRV is non-ideal, PAL captures activity context more frequently, i.e. 3-minute intervals. Users can customize all application parameters, e.g. HRV range, recording frequency, etc.

**DISCUSSION AND FUTURE WORK**

We plan to test and deploy PAL real-time, personalized and context-aware health and cognition support system in these ways: i. Deploy PAL with different people to get long-term data about their activities and physiological states, and evaluate the usefulness and accuracy of that data for users; ii. Evaluate the efficacy of user's self-support interventions and study the best types and times of interventions; iii. Work with psychologists and therapists to evaluate different personalized context-aware interventions for supporting mindfulness, habit change, growth mindset, learned optimism, stress management, etc; iv. Collaborate with researchers to evaluate our cognitive augmentation use cases, e.g. language learning and memory augmentation. We also plan to technically improve PAL's platform in these ways: i. Add more sensors to PAL, e.g. biochemical sensors or electroencephalogram (EEG), to track and correlate more of user's physiological states with their activities; ii. Test and add more feedback modalities, e.g. smell, haptics, visual cell phone notifications, etc; iii. Improve our machine learning and data visualization for PAL's data, e.g. to predict different user contexts, e.g. stress, anxiety, and panic attacks, to preemptively intervene even before the user gets in undesired states; iv. Learn the most effective personalized behavior change interventions for each user, possibly using reinforcement learning; v. Deploy more on-device machine learning models, e.g. for speech processing, human activity recognition, and food tracking, for more real-time interventions; iv. Create an open data platform for sharing PAL's data safely, e.g. by using differential privacy.

**CONCLUSION**

Self-improvement is a booming industry as people want to improve themselves. However, most self-improvement resources, e.g. books and therapists, are not always present with the user and may not even have the detailed information required to help users in personalized ways. We designed PAL to be a user's constant companion to help users track their activities and physiological states over time tandem and learn the correlations between their activities and physiological states. PAL's context-aware platform also provides real-time, personalized and context-aware interventions to users to not only empower self-awareness, but also self-help and self-change. We have PAL's platform open-source so that users, developers, researchers, doctors, and therapists can use it in extensible and modular ways for real-time, personalized and context-aware "If This Then That (IFTTT)" applications for health and cognition. We will continue to test, develop and deploy PAL for personalized and participatory health and cognition support.